\documentclass[epjCONF]{svjour}
\newcommand{\order}[1]{ \mathcal{O} \left( #1 \right) }
\newcommand{\ave}[1]{\left\langle #1 \right\rangle}

\usepackage{graphics,epsfig}
\usepackage[varg]{txfonts} 
\usepackage[latin1]{inputenc}
\session-title{Resonances workshop, Austin, Texas, march 2012}
\begin{document}
\title{Resonances and fluctuations in the statistical model}
\author{Giorgio Torrieri\thanks{\email{torrieri@th.physik.uni-frankfurt.de}}}
\institute{$^a$FIAS,
  J.W. Goethe Universit\"at, Frankfurt A.M., Germany }
\abstract{
We describe how the study of resonances and fluctuations can help constrain the thermal and chemical freezeout properties of the fireball created in heavy ion collisions.    This review is based on \cite{hornreso,ratio1,ratio2,prcfluct,westfall}
} 
\maketitle
\section{Introduction}
The idea of modeling the abundance of hadrons using statistical mechanics techniques has a long and distinguished history ~\cite{history1,history2,history3,history4,jansbook}.  In a sense, any discussion of the thermodynamic properties of hadronic matter (e.g. the existence of a phase transition) {\em requires} that statistical mechanics be applicable to this system ( through not necessarily at the freeze-out stage).

That such a model can describe {\em quantitatively} the yield of most particles, including multi-strange ones, has in fact been indicated by fits to average particle abundances at AGS,SPS and RHIC energies  ~\cite{jansbook,bdm,equil_energy,becattini,nuxu,share,castorina}.   The jump from {\em noting} the qualitative goodness of this description to ascertaining its limits, and using it as a tool to study dense hadronic matter, is however a still ongoing process.

  Some practitioners \cite{history1} interpret the statistical model results in terms of nothing more than phase space dominance:  For a process strongly enough interacting with enough particles in the final state, dynamics ``factors out'' into a normalization constant, and the final state probabilities are dominated by phase space.  If this is the case, the applicability of the statistical model has nothing to do with a genuine equilibration of the system.
Others think that in soft QCD processes particles are ``born in equilibrium'' \cite{castorina}, and the applicability of the statistical model to even smaller systems  is a fundamental characteristic of QCD.
Still others \cite{bdm,jaki} believe that the applicability of the statistical model is a sign of a phase transition, as the chemical equilibration of hadrons signals a regime in which multi-particle processes and high-lying resonances dominate.

Obviously, the link between the QCD phase diagram, defined in the Grand Canonical limit, and experimental data can only be made in the last interpretation.    Yet it is not clear how these interpretations can be differentiated.    Currently, the debate centers around the scope of application of the statistical model \cite{castorina,pbmel} to smaller systems.
Yet in every one of these cases an objective criterion for linking the {\em goodness of fit} to {\em the effective relevance of the underlying theory} is still absent \cite{castorina}:  Since it is clear that,in all regimes, {\em non-statistical processes} such as jet fragmentation and ``corona physics'' are also present, and since error bars vary dramatically across system sizes for essentially experimental reasons, excluding the statistical model from a raw $\chi^2$ analysis is highly nontrivial \cite{castorina,pbmel}.  

Two, related, observables which might lead to progress in this context are short-lived QCD resonances \cite{ratio1,ratio2} and event-by-event fluctuations \cite{prcfluct,sharev2}.

Short-lived resonances go to the heart of this question because,in principle, their abundance can be modified by {\em elastic interactions } of already formed hadrons.     The abundance of hadrons stable against strong decays, such as the $\Lambda$, is expected to be  unchanged even if particles created at hadronization continue to interact, because inelastic interactions (such as $\Lambda \pi \rightarrow p K$) are suppressed w.r.t. purely elastic scattering.   Even purely elastic scattering, such as $\Lambda \pi \rightarrow \Sigma^* \rightarrow \Lambda \pi$ could in principle create new $\Sigma^*$s or make $\Sigma^*$s created at hadronization undetectable (since such particles are detectable by invariant mass reconstruction only).    Since the fitted hadronization temperature of $T=170 $ MeV generally implies a non-negligible reinteraction phase for A-A collisions, but not for smaller systems, the applicability of the statistical model to describe unstable resonances as a function of system size is a good gauge to see to what extent the statistical model applies specifically {\em at the hadronization stage}.

As for fluctuations, it is a fundamental principle of statistical mechanics that variances around averages scale w.r.t. averages in a way defined by the maximization of entropy under the constraints specific to the ensemble.  
In our context ``Averages'' are particle multiplicities per event and fluctuations are event-by-event fluctuations.   For macroscopic systems, this principle ensures that fluctuations become negligible and the expectation that the state of the system is the maximum entropy one is nearly certain to be realized.
$\order{100-1000}$ particles is not enough for this to be the case, but, if statistical mechanics applies, one should still see that yields, fluctuations and higher cumulants scale in a way calculable from the partition function.   

These two classes of observables are actually related \cite{westfall,jeon,shuryak}: Even {\em invisible} resonances in the medium, whose decay products are rescattered and thus undetectable by invariant mass reconstruction still continue to ``exist'' as a correlation between their decay products abundances.    In contrast, ``regenerated'' resonances do not change such correlations 
, because they are formed by particles already generated at hadronization.
Comparing resonance abundances with fluctuation observables can be used to gauge, from experimental observables, whether hadronization is where particle abundances are fixed once and for all, or whether further dynamical evolution also impacts such abundances.

In the statistical model there are two types of chemical equilibrium~\cite{jansbook}.:
all models assume relative chemical equilibrium, but some 
also assume absolute chemical equilibrium, implying the presence of  the ``right'' abundances of valance  up, down, and strange quark pairs to be present after hadronization.   In absolute 
chemical equilibrium at highest heavy ion reaction energy  
one obtains chemical freeze-out temperature   $T \sim 160-170$ MeV, which goes down to $\sim 50$ MeV at the lowest reaction energies \cite{equil_energy}.

\begin{figure} 
\epsfig{width=7cm,clip=,figure=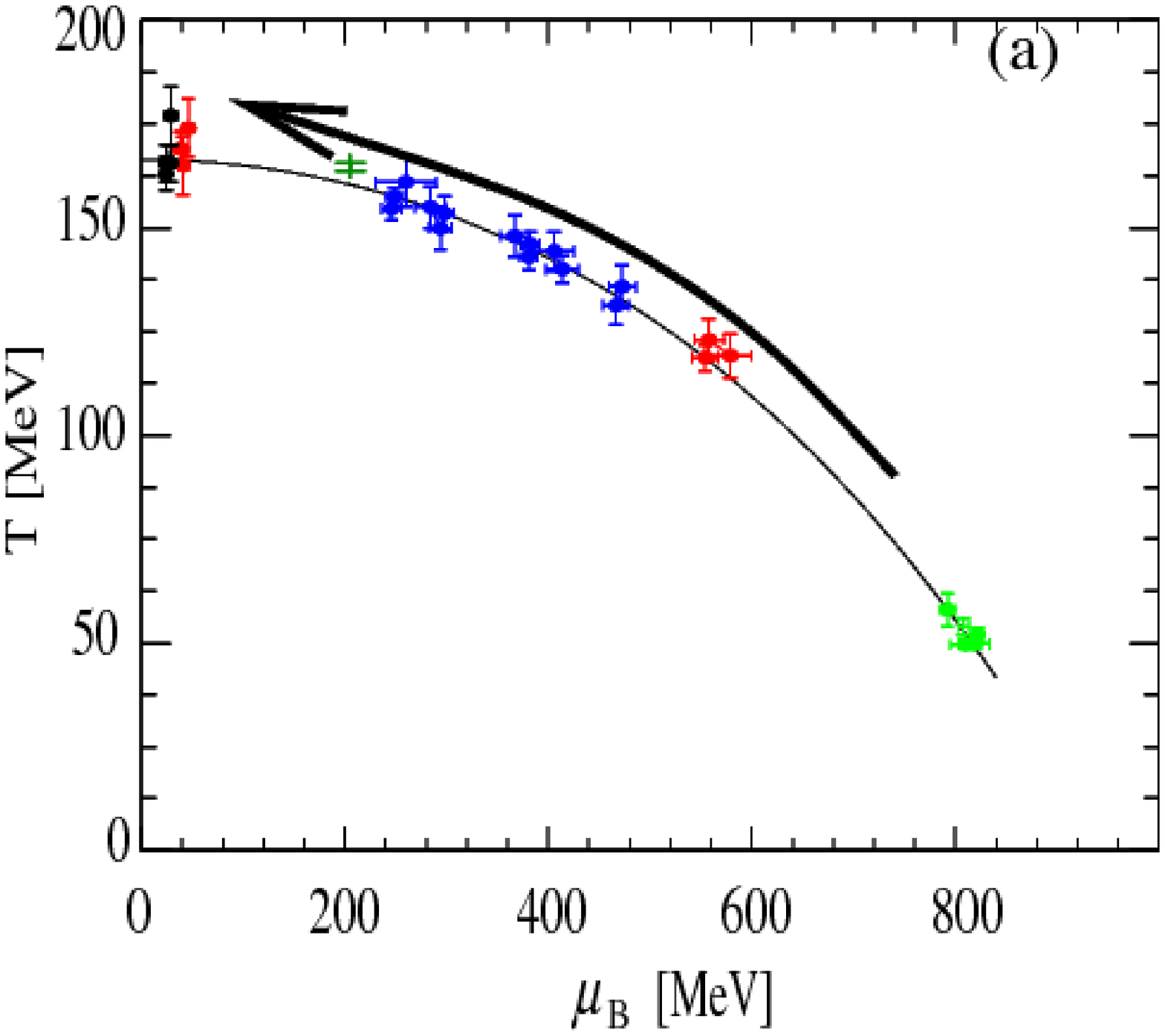}
\epsfig{width=7cm,clip=,figure=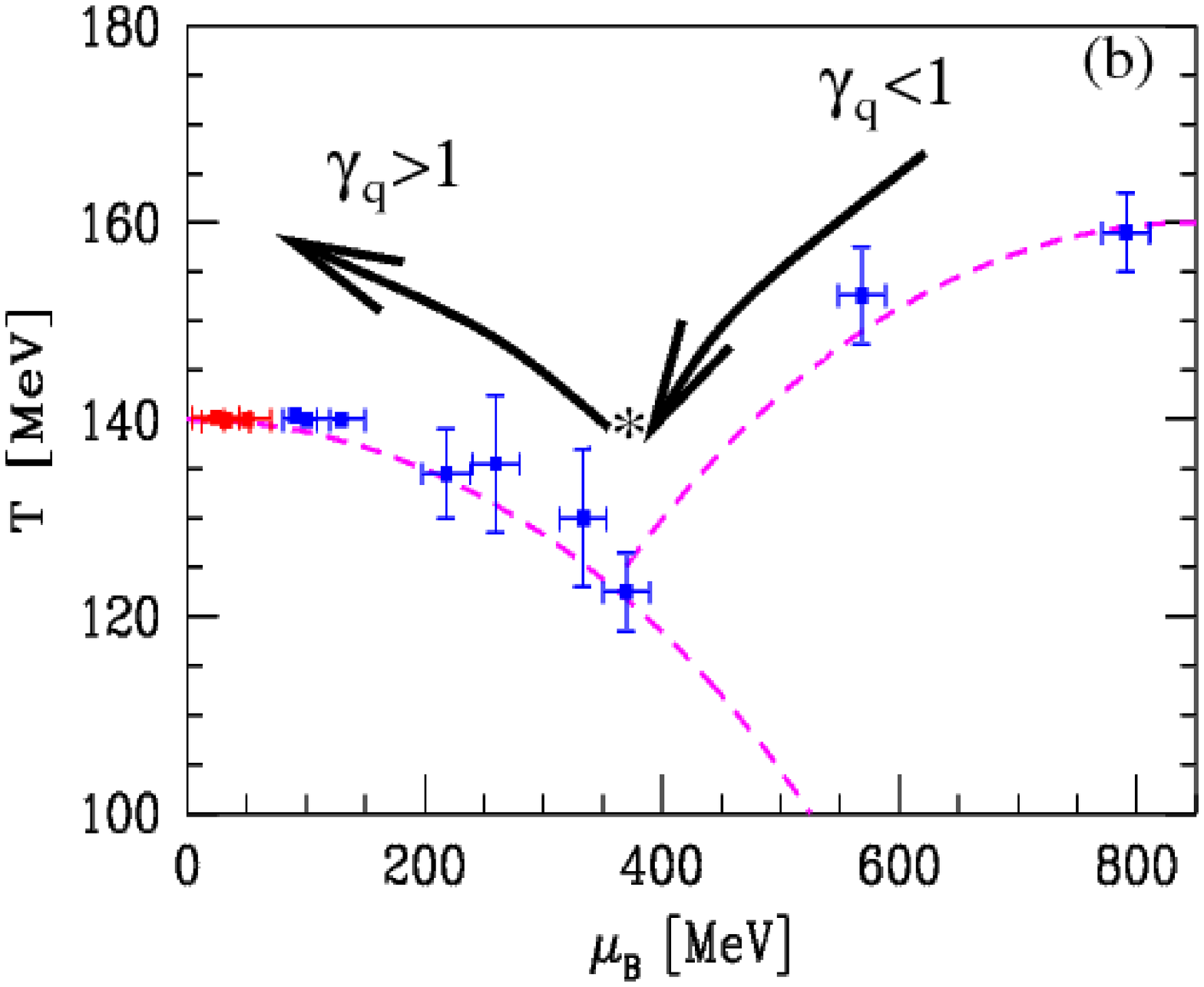}
\vskip -0.5cm
\caption{(Color online)\label{phaseall} 
Dependence of freeze-out temperature  $T$ and baryo-chemical potential $\mu_{\rm B}$
on reaction energy in the Equilibrium 
(panel (a), \cite{equil_energy}) and non-equilibrium (panel (b),\cite{gammaq_energy}) 
freeze-out.  The arrow corresponds to increasing
$\sqrt{s}$.  }
\end{figure}

The energy dependence of the freeze-out temperature \cite{hornreso} than follows the
trend indicated in panel (a) of figure \ref{phaseall}: as the collision energy increases, the freeze-out temperature 
increases and the baryonic density (here baryonic chemical potential $\mu_{\rm B}$)
decreases~\cite{equil_energy}.  An  increase of freeze-out temperature with $\sqrt{s}$ is 
expected on general grounds, since  with increasing reaction energy a 
greater fraction of the energy is carried by mesons created in the collision, 
rather than pre-existing baryons~\cite{hornreso}. 

Further refinements  in the approach  described above
are often implemented, the more notable ones are
allowance for strangeness chemical nonequilibrium \cite{becattini} at low $\sqrt{s}$ and canonical effects for small reaction volumes \cite{canonical1}.
These effects do not materially alter the behavior of
temperature and chemical potential shown in the panel (a) of
Fig. \ref{phaseall}.    Such smooth variations fail to fully reproduce the non-continuous  features in the energy dependence of hadronic observables, 
such as the ``kink'' in the multiplicity per number of participants and the 
``horn''~\cite{equil_energy,horn,gammaq_energy} in certain particle yield ratios.   

Non-monotonic behavior of particle yield ratios could  indicate
a novel reaction mechanism, e.g. onset of the deconfinement 
phase~\cite{horn}.  In a rapidly expanding fireball undergoing a change in microscopic degrees of freedom, it is a possibility
that absolute chemical equilibrium does not hold.   Either super-cooling \cite{jansbook} or bulk viscosity-triggered instabilities \cite{visc1,visc2} could justify such lack of chemical equilibrium on microscopic grounds.

One can model this phenomenologically by removing the hypothesis of absolute chemical equilibrium among hadrons 
produced. The   systematic behavior of 
$T$ with energy in this case is quite different~\cite{gammaq_energy},
as  is shown  in panel (b) of
figure \ref{phaseall}.   
The two higher $T$ values  at right are for 20 (lowest SPS) and  (most to right) 11.6 $A$ GeV (highest AGS)
reactions. In these two cases the source of particles is  a hot chemically under-saturated  ($T \sim 170$ MeV ) fireball.
Such a system could be a conventional hadron gas fireball that 
had not the time to chemically equilibrate. 

At higher heavy ion reaction energies it is possible \cite{jansbook} to match the entropy of the emerging 
hadrons with that of a system of nearly massless partons when one considers supercooling
to $T\sim 140$ MeV (essentially bringing chemical and thermal freeze-out close together, as seen in \cite{jan_search,flork}),  while both light and strange quark phase space in the hadron stage 
acquire significant over-saturation with the phase space occupancy 
$\gamma_{q=u,d}>1$ and at higher energy also $\gamma_s>1$.  A drastic change 
 in the non-equilibrium condition occurs near 30 $A$ GeV,
corresponding to the dip point on right in panel (b) of the figure \ref{phaseall} (marked by an asterisk). 
At heavy ion reaction energy below (i.e. to right in panel (b) of
figure \ref{phaseall}) of this point, hadrons have not reached 
chemical equilibrium, while   at this point, as well as, at heavy ion reaction energy above 
(i.e. at and to left in panel (b) of figure \ref{phaseall}),   hadrons  emerge from a much denser and chemically more saturated system,
as would be expected were QGP formed at and above 30 $A$ GeV.   
This is also  the heavy ion reaction energy corresponding to the ``kink'', which tracks the QGP's entropy density 
(higher w.r.t. a hadron gas), and the peak of the ``horn'' \cite{horn}, 
which tracks the strangeness over entropy ratio (also higher w.r.t. a hadron gas). 

 The main reason for the wider acceptance of the equilibrium 
approach $\gamma_i=1$  is its greater simplicity,
there are fewer parameters. Moreover, considering the quality of 
the data the non-equilibrium parameter $\gamma_q$ is not 
necessary to pull the statistical significance above  it's 
generally accepted  minimal value of 5 $\%$.  On the other hand,
the   parameters $\gamma_q$ and $\gamma_s$ were introduced on 
{\em  physical} grounds \cite{jansbook}, thus 
these are not   arbitrary fit parameters.   Moreover,  these
parameters, when used in a statistical hadronization  fit, 
converge to theoretically motivated values. They also  
help to explain the trends observed in the energy dependence 
of hadronic observables. 
\section{Resonances}
 Many strong interaction resonances, a set we denote by the collective symbol $N^*$ (such as 
$K^{*0}(892), $ $ \Delta(1232), $ $\Sigma^* (1385), $ $\Lambda^* (1520), $ $\Xi^*$ (1530)) 
carry the same valance quark content  as their ground-state 
counter-parts $N$ (corresponding: $K,$ $p,$ $\Sigma,$ $\Lambda,$ $\Xi$). $N^*$
 typically decay by emission of a pion,  $ N^* \rightarrow N+\pi$.   
Considering the  particle yield ratio $\ave{N^*}/\ave{N}$ in the Boltzmann approximation (appropriate for the particles considered), 
we see that all chemical conditions and parameters (equilibrium and non-equilibrium) cancel out, and the 
ratio of yields between the   resonance and it's
ground state is a function of the masses, and the freeze-out temperature, with 
second order effects coming from the cascading decays of other, more massive 
resonances~\cite{jansbook,share}:
\begin{equation}\label{relY}
\frac{\ave{N^*}}{\ave{N}} \simeq \frac{g_{N^*} W \left(\frac{ m_{N^*}}{T}\right)
         +\sum_{j \rightarrow N^*}  b_{jN^*}\,g_{j} W \left(\frac{ m_{j}}{T}\right)}
              { g_{R} W \left(\frac{ m_{R}}{T}\right)
        + \sum_{k\rightarrow R}  b_{kR}\, g_{k}W \left(\frac{ m_{k}}{T}\right)}
\label{ratio}
\end{equation}
where $W(x)=x^2K_2(x)$ is the (relativistic) reduced one particle  phase space, $K_2(x)$
being a Bessel function, 
$g$ is the quantum degeneracy, and $b_{jR}$ is the branching ratio of resonance $j$ decaying into $R$.

Because of the radically different energy dependence of freeze-out temperature 
in the scenarios of \cite{equil_energy} and \cite{gammaq_energy}, seen in figure  \ref{phaseall},
the prediction for the resonance ratios Eq. (\ref{relY})  vary greatly between these two scenarios.  
In the equilibrium scenario the temperature goes {\em up} with heavy ion reaction energy, and thus 
 the resonance abundance should go smoothly up for all resonances.
On the other hand, the nonequilibrium scenario, with a low temperature arising only in some 
limited reaction energy domain, will lead to   resonance abundance which
should  have a clear dip  at that point, but otherwise remain relatively large. 

\begin{figure} 
\epsfig{width=7.1cm,clip=,figure=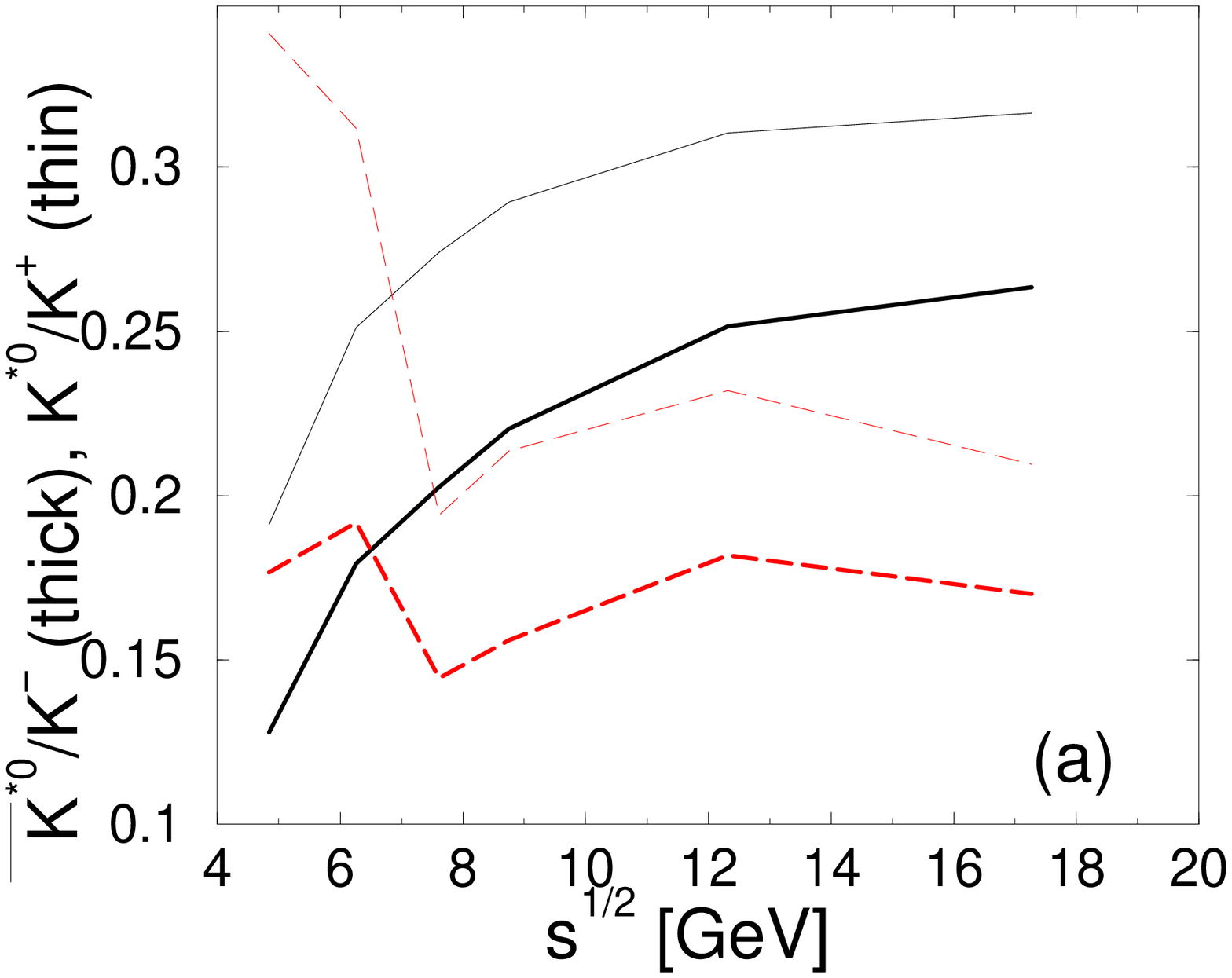}
\epsfig{width=7.1cm,clip=,figure=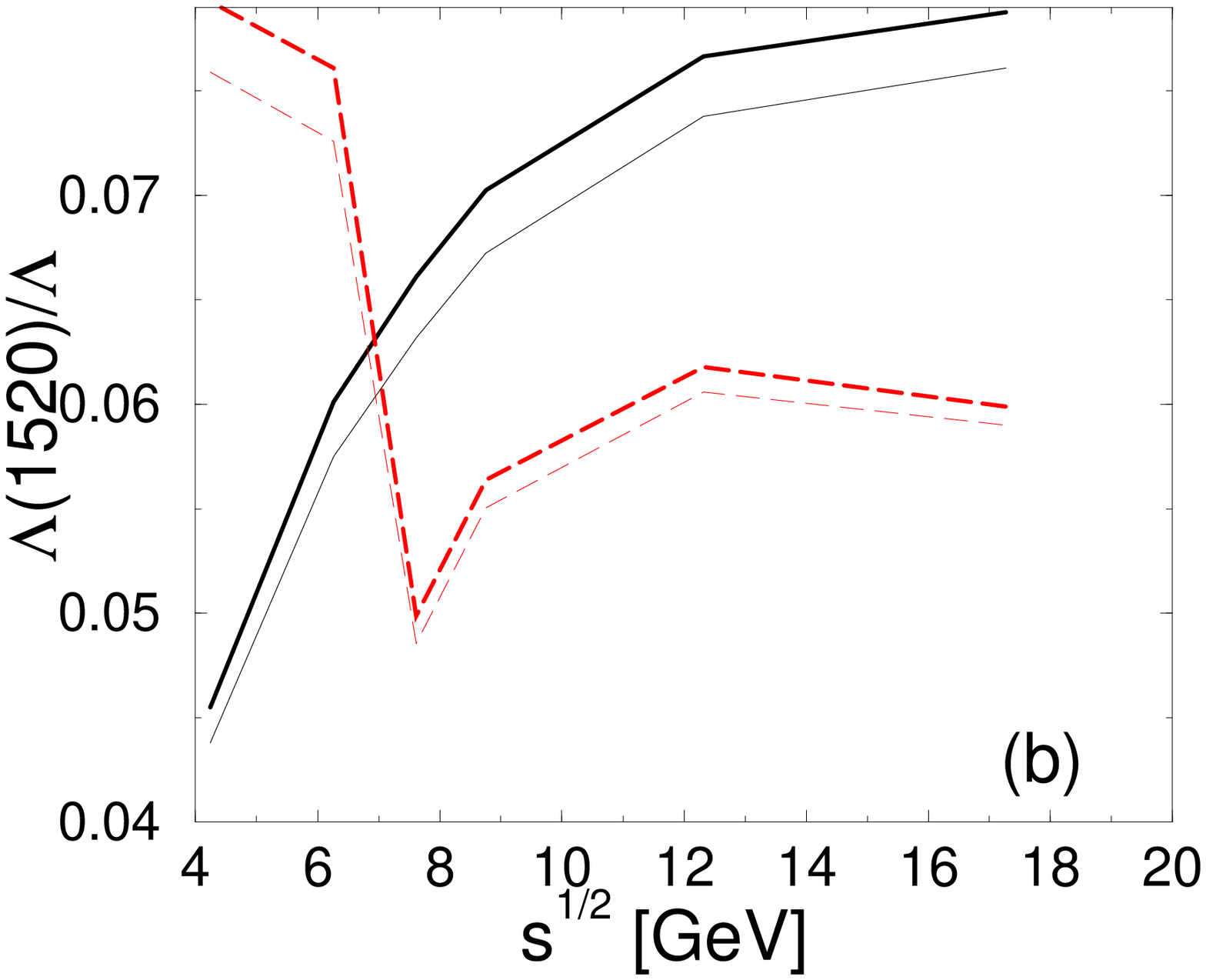}
\epsfig{width=7.1cm,clip=,figure=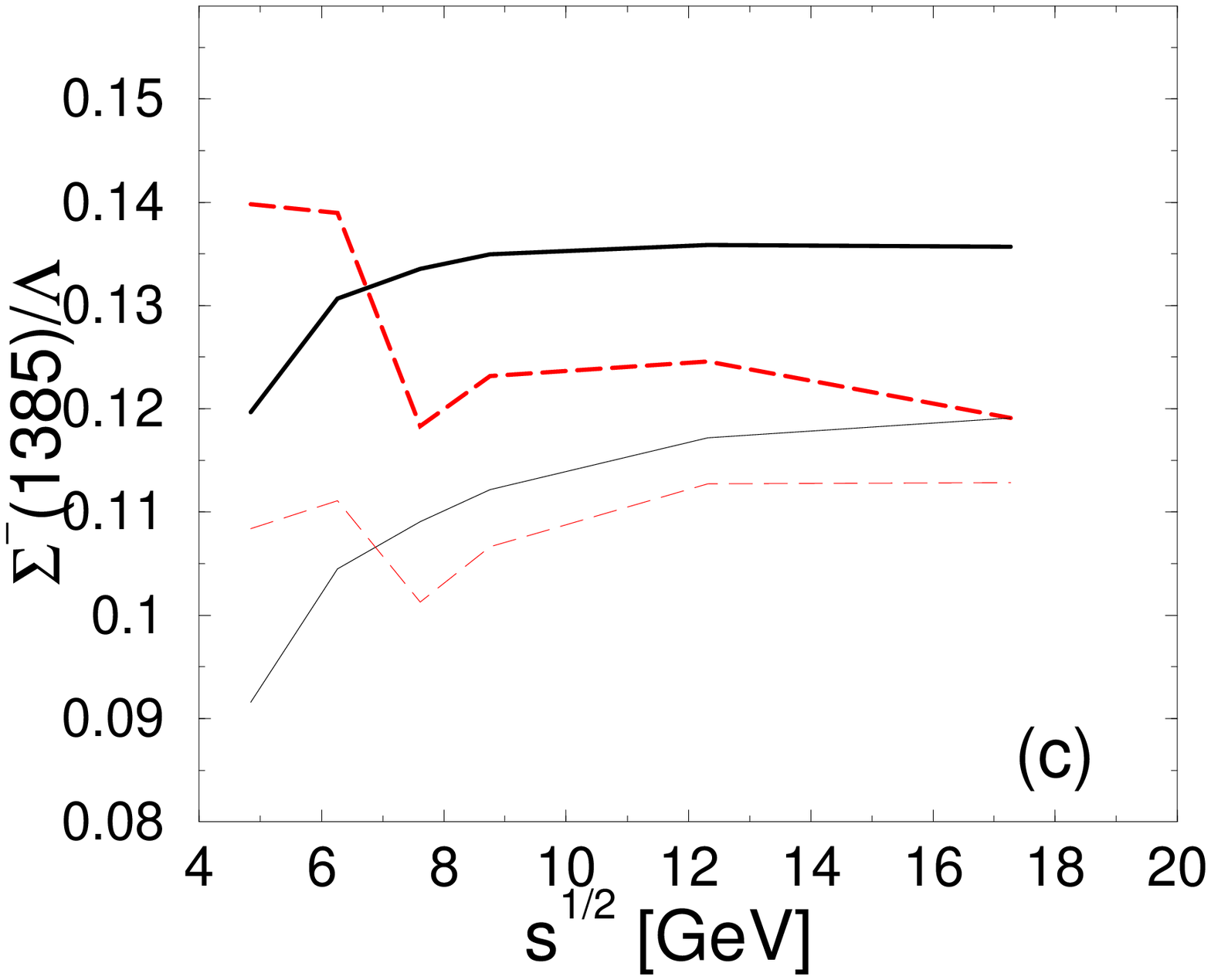}
\caption{(Color online)\label{figres}
Ratio of resonance to the stable particle. Thick lines for particles with strange quark content, 
thin lines for particles with anti-strange quark content,  as a function of energy.
Solid black lines refer to the equilibrium fits ($\gamma_{q,s}=1$), with the parameters for AGS
and SPS energies taken from \cite{equil_energy}.  Dashed red lines refer to
non-equilibrium  fits ($\gamma_{q,s}$ fitted), with the best fit parameters for 
AGS and SPS energies taken from \cite{gammaq_energy}.  }
\end{figure}

We have evaluated several resonance relative ratios shown in  figure \ref{figres} within the two scenarios, 
using the   statistical hadronization code SHARE \cite{share,sharev2}.
For the non-equilibrium scenario, we have used the parameters given in \cite{gammaq_energy}, 
table I.   For the equilibrium scenario, we used the parametrization given in \cite{equil_energy}  
figures 3 and 4.
In the latter case, the strangeness and isospin chemical potentials  
were obtained by requiring net strangeness to be zero, and net charge per baryon to be  the same as in the colliding system.  
\begin{figure}
\epsfig{width=7.1cm,clip=,figure=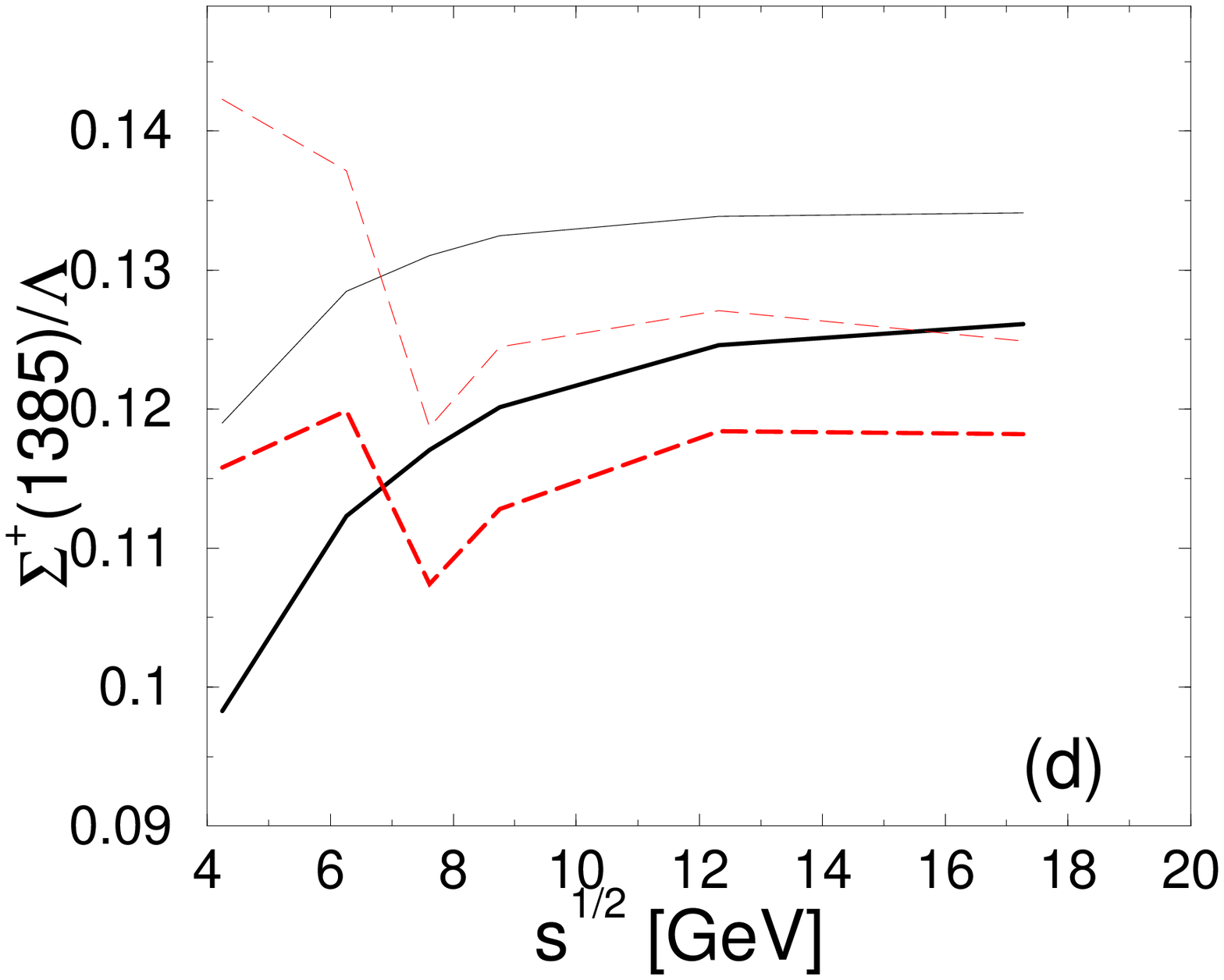}
\epsfig{width=7.1cm,clip=,figure=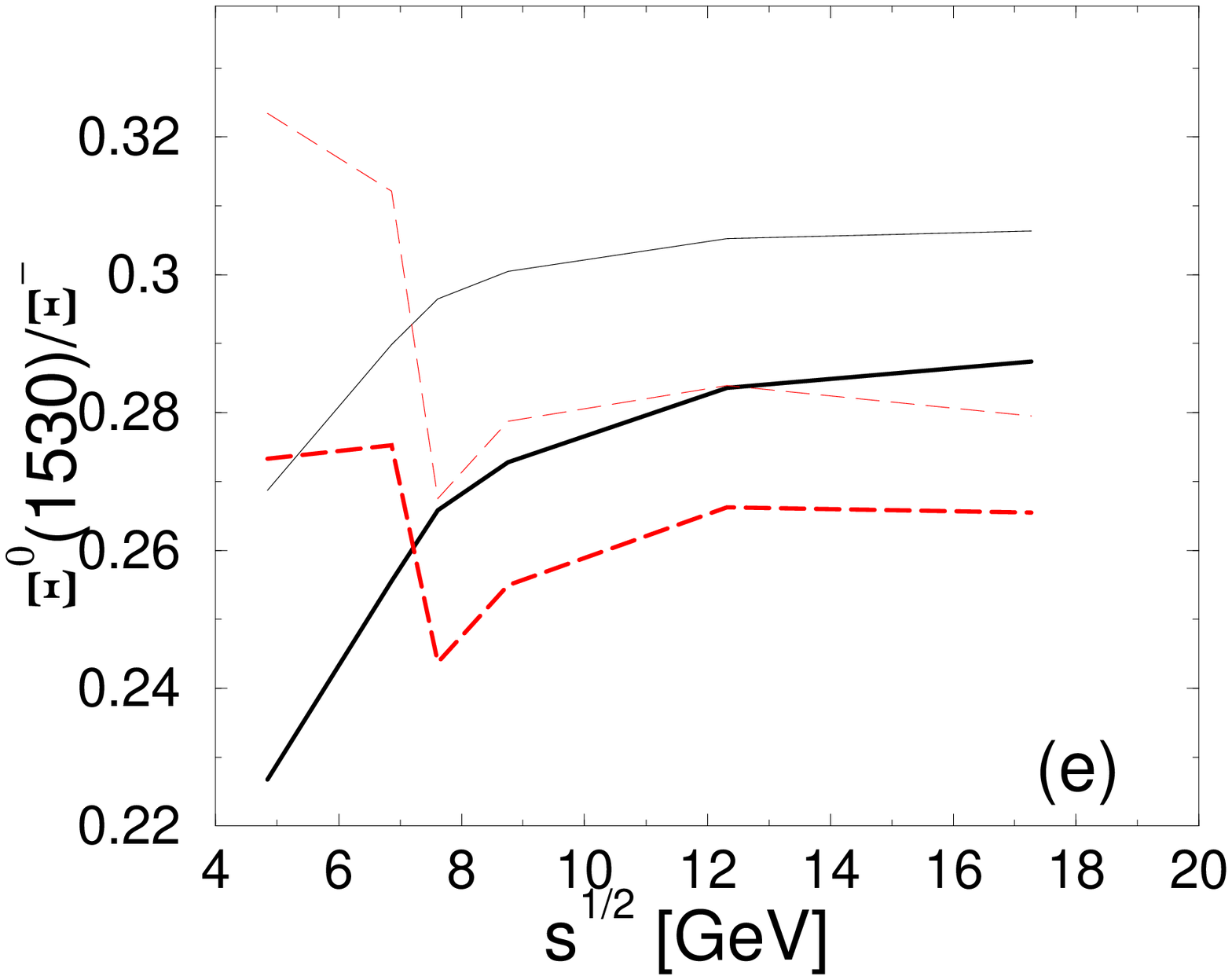}
\epsfig{width=7.1cm,clip=,figure=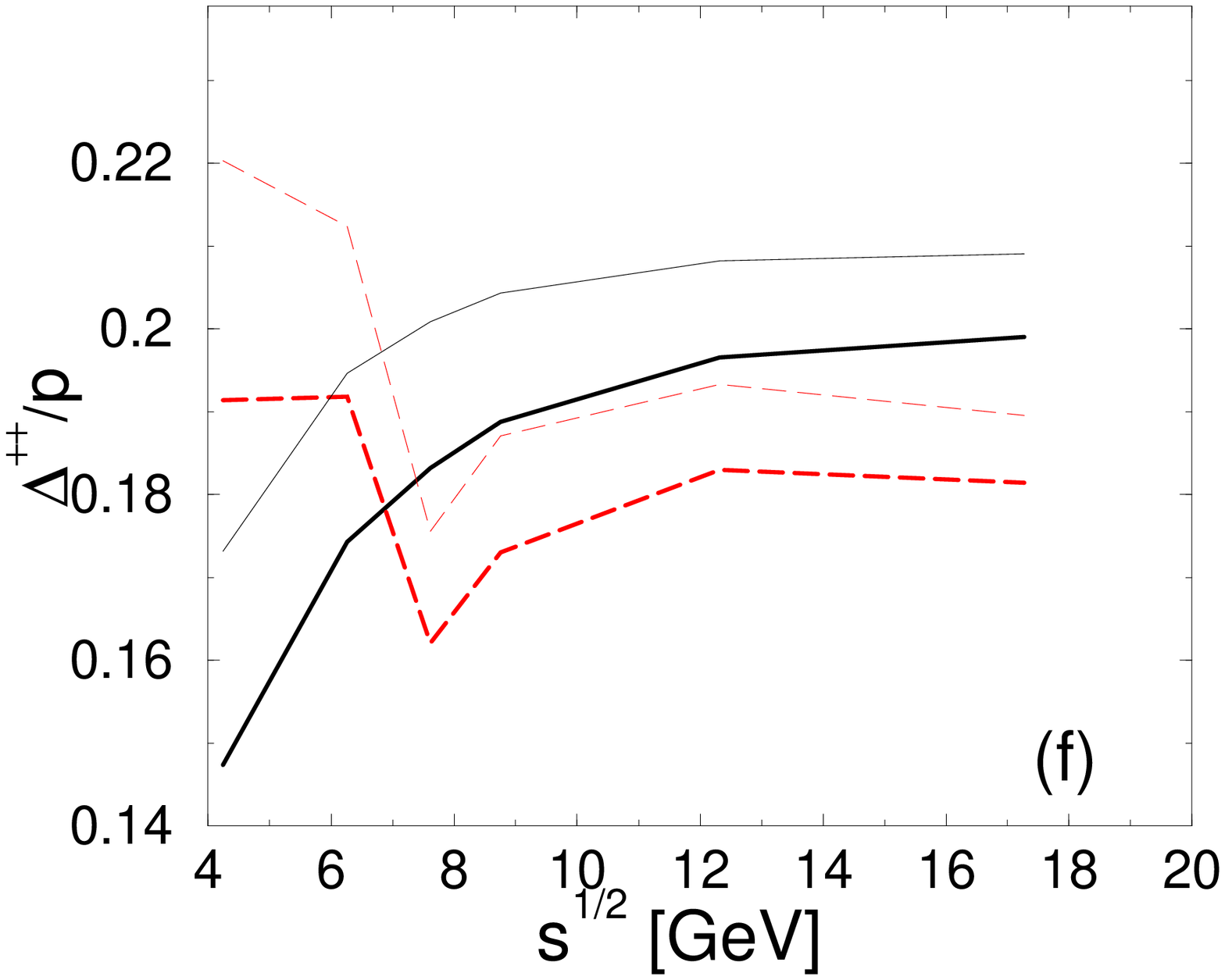}
\caption{\label{figres2} Same as Fig. \ref{figres},more resonances} 
\end{figure}

 As  seen    in figure \ref{figres}, the expected trend with $\sqrt{s}$   
is apparent in all  considered resonance ratios, though in
cases where the mass difference is large,
the effect is much more pronounced than in some others. Indeed, 
for many of the ratios we present the experimental error may 
limit the usefulness of our results
However, because of the cancellation (to a good approximation) of the baryo- 
and strangeness chemical  potentials, the qualitative
prediction for the {\em energy dependence} of the resonance yields within 
the two models is robust. Namely, 
within the chemical equilibrium model the 
temperature of chemical freeze-out must steadily increase and so
does the $\ave{N^*}/\ave{N}$ ratio. For the chemical non-equilibrium 
model the $\ave{N^*}/\ave{N}$ dip primarily relies on the response of $T$   
to the degree of chemical equilibration: prior to chemical equilibrium   
 for the valance quark abundance, at a relatively low reaction energy, the 
freeze-out temperature 
$T$ is relatively high.   At a critical energy, $T$ drops as  the 
hadron yields move to or even exceed light quark chemical equilibrium,
yet reaction energy is still not too large, and thus the baryon density is high
and meson yield low.
As reaction energy increases further, 
$T$ increases  and the $\ave{N^*}/\ave{N}$ yield from that point on increases. 
The drop in $\ave{N^*}/\ave{N}$ at critical $T$,
would be completely counter-intuitive in an equilibrium picture.
It would hence provide evidence that non-equilibrium
effects such as supercooling, where such a drop would be expected, 
are at play.

One difficulty of this approach is that, unlike for stable particles, pseudo-elastic processes such as
$N \pi \rightarrow N^* \rightarrow N\pi $ 
and post-decay $ N^* \rightarrow N\pi $
scattering of decay products in matter 
could potentially considerably alter the observable  final ratio of {\em detectable} $N^*$ to $N$.   
The combined effect of rescattering and regeneration has not been well understood.  
We have argued that the formation  of additional 
{\em detectable} resonances is negligible~\cite{ratio1,ratio2}, while scattering
of decay products can decrease the visible resonance yields except for 
sudden hadronization case.   Transport simulations \cite{urqmdreso} confirm that rescattering generally dominates over regeneration in a hadronic medium.
This is not surprising:
If the number of reinteractions is small, rescattering generally dominates.  If the number of reinteractions is 
large, detailed balance would mean $K^*/K$ would reequilibrate from the higher chemical freeze-out to the lower thermal freeze-out temperature.

To obtain a further quantitative estimate, we then evolved in time the ratios using a model which combines an average 
rescattering cross-section with dilution due to a constant collective expansion.
In this model, the initial resonances decay with width $\Gamma$ through 
the process $N^* \rightarrow N_D \pi$.  
Their decay products $\ave{N_D}$ ($\ave{N_D}(t=0)=0$) then undergo rescattering at a rate
proportional to the medium's density as well as the average rescattering rate.
\begin{eqnarray}
\label{rescattering}
 \frac{d \ave{N^*}}{d t}& =& -\Gamma \ave{N^*}, \\
 \frac{d \ave{N_D}}{d t}&=& \Gamma \ave{N^*} -D\sum_j
\langle \sigma_{Dj} v_{Dj}\rangle \rho_j
          \left(\frac{R_0}{R_0+v t}\right)^3. 
\end{eqnarray}
$v$ is the expansion velocity, $R_0$ is the hadronization radius, 
$\rho_j=  n_{j} (m_j,T)$ the initial hadron density
and $\langle \sigma_{Dj} v_{Dj}\rangle$ is the  hadron average
flow and interaction cross-section.

\begin{figure}
\psfig{width=7.cm,clip=1,figure= 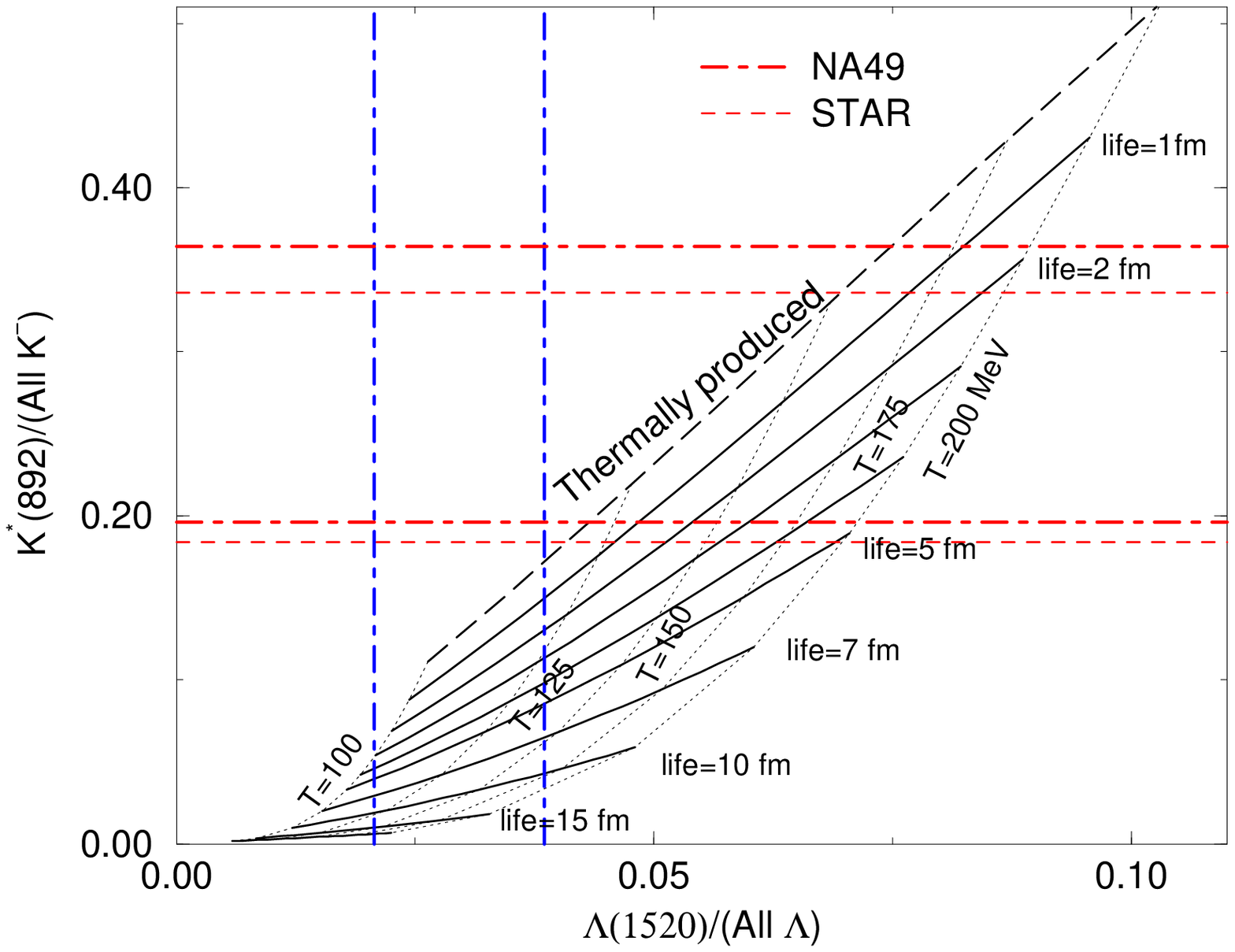}
\psfig{width=7.cm,clip=1,figure= 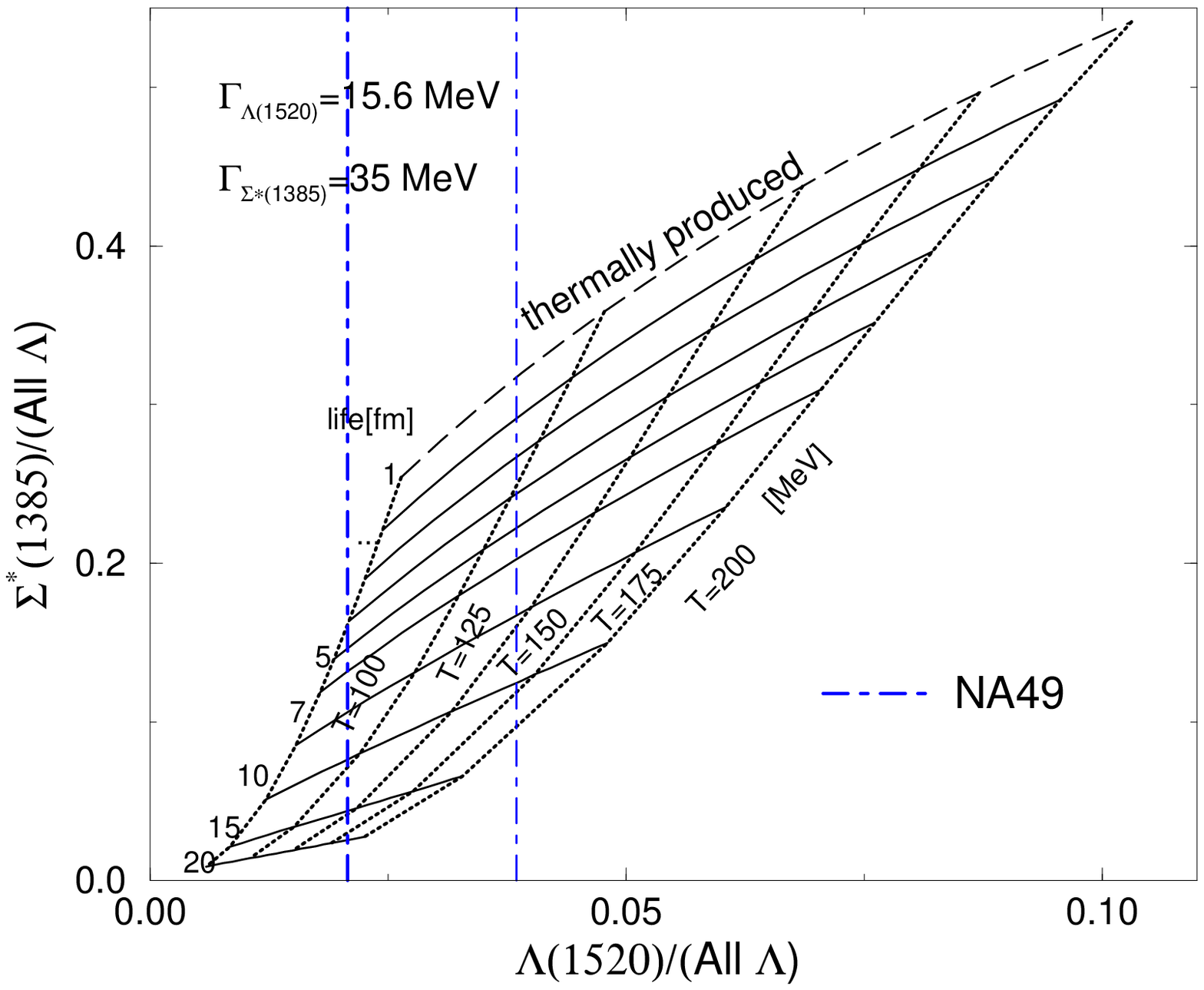 }
\psfig{width=7.cm,clip=1,figure= 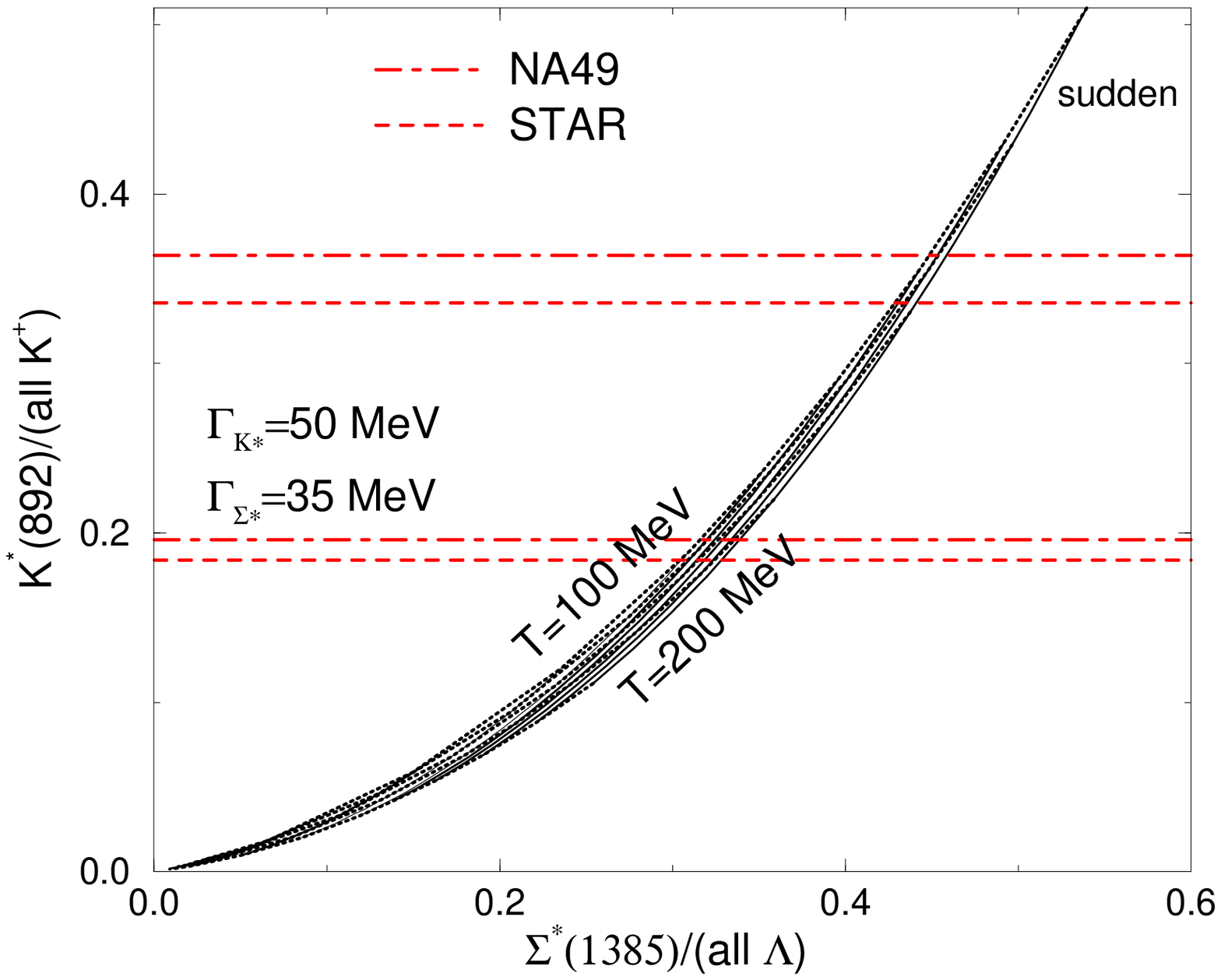 }
\caption{Two $\ave{N^*}/\ave{N}$ diagrams over
a mesh of temperature and lifespan.
 Straight lines give  experimental results \cite{fluctreso,na49reso}.
 \label{projdiag}}
\end{figure}

In this calculation, we have neglected the regeneration term 
$R\propto \langle \sigma_{Di}^{ INEL} v_{Di}\rangle  \rho_{i}$, 
since  detectable regenerated resonances
need to be real (close to mass-shell) particles.
Figure \ref{projdiag} shows how the ratios of 
$\Sigma^*/\Lambda$ and $K^*/K$ evolve
with varying hadronization time within this model.
It is apparent that measuring two such 
ratios simultaneously gives both the hadronization 
temperature and the time during which rescattering
is a significant effect.
Figure \ref{projdiag} shows the application of this 
method to $K^*,\Lambda(1520)$ and $\Sigma^*$ \cite{fluctreso,na49reso}.

 Assuming a long lived hadron phase, the
energy dependence of most of the resonance ratios considered here has been calculated in a
hadronic quantum molecular dynamics model. The result   (figure 7 and 8 in~\cite{urqmdreso})
is qualitatively similar to the chemical equilibrium results for resonance ratios, we see a smooth rise with energy.
Thus, in the case of chemical equilibrium, with a considerable separation between chemical and thermal freeze-out
inherent in Ref.~\cite{urqmdreso}
rescattering and regeneration will affect the {\em quantitative} $\ave{N^*}/\ave{N}$ ratio, but will not alter
the {\em dependence on heavy ion reaction energy} shown in figure~\ref{figres}.

Thus, while at a given energy the role of post-freezeout rescattering/regeneration is still somewhat ambigous, an energy scan (experimentally conducted or planned around several labs \cite{horn,fair,rhicscan,nica}) of resonance production should ameliorate some of this ambiguity.   In the next section we will explain how
fluctuations could make this ambiguity be experimentally resolvable at each given $\sqrt{s}$ too.
\section{Fluctuations}
Before we discuss the significance of fluctuations in a statistical model, we need to discuss some  issues relevant to fluctuations and higher cumulants of particle distributions.  Generic fluctuation observables are non-trivially dependent both on {\em non-statistical} fluctuations in the systems volume, which can not be described by the statistical model, and on acceptance limitations of the detector.

Regarding the first issue, due to our incomplete understanding of how the initial state and dynamics contribute to fluctuations at freeze-out, the best approach is to choose an observable which is {\em insensitive} to any volume fluctuations.   In the thermodynamic limit, where volume becomes a proportionality constant at the level of the partition function, a tempting observable is the scaled variance of the {\em ratio} of two particle multiplicities measured event by event.
That this is in fact a good guess, and all dependence of volume factorizes, can be proven in the thermodynamic limit to order $\sim \ave{(\Delta V)/\ave{V}^2}$ \cite{westfall}.
The residual dependence of $\sigma_{N_1/N_2}$ on the average volume $\ave{V}$ can be in turn eliminated, in the grand canonical ensemble, by focusing on $\Psi=\ave{N_1} 
\sigma^{N_1/N_2}_{dyn}$, where $\ave{N_1}$ and $\sigma_{dyn}$ are to be {\em measured within the same acceptance}.  Note that this independence is specific to the grand-canonical ensemble, so should not apply to scenarios where the ``enhancement of strangeness'' in A-A collisions is due to the transition between the canonical limit in elementary processes and the grand canonical limit in A-A \cite{canonical1}.   In this scenario, the ``strangeness correlation volume'' should regulate $\Psi$ as in  \cite{canfluct}.

A different problem is the effect of a detectors limited acceptance
( Particle (mis)identification, Limited rapidity and momentum resolution, momentum cuts necessary to eliminate jets etc) on fluctuation observables.  These are much more difficult to model than averages, and once again an observable needs to be constructed insensitive to them.
Hence, the necessity of mixed event subtraction \cite{methods}.   Mixed events here, are {\em defined} as events where {\em no physical correlation from the original event} are left in.  This means that any correlation seen is due to the imperfection of the detector (the fact that detector acceptance excludes particles of $p_T>1$ GeV for both event A and B creates a correlation between $A$ and $B$).
To a good approximation, examined in the next paragraph, the measured fluctuation is $\sigma^2 = \sigma_{physics}^2 + \sigma_{acceptance}^2$
and the mixed event one is $\sigma_{mix}^2 = \sigma_{trivial}^2 + \sigma_{acceptance}^2$.
Therefore, concentrating on $\sigma_{dyn}^2 = \sigma^2 - \sigma_{mix}^2$
\textit{should} eliminate acceptance effects \cite{methods}.  
\begin{figure}[h]
\begin{center}
\epsfig{width=8cm,figure=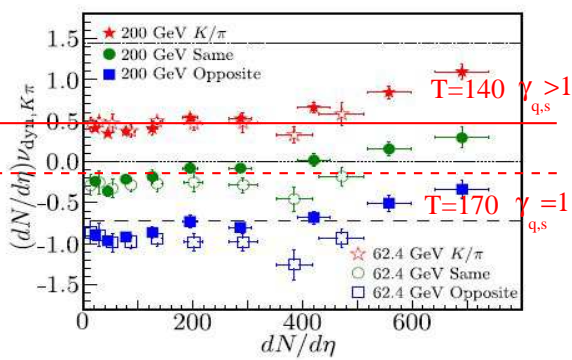}
\caption{\label{fluctstar}    Scaling of $K/\pi$ fluctuations at RHIC 200 GeV \cite{fluctstar}. }
\end{center}
\end{figure}

  Particle abundances and fluctuations can be calculated from the first and second derivatives a particle's partition function, and related to the fluctuation of a ratio via \cite{sharev2}
\begin{equation}
\sigma_{N_1/N_2}^2 = \frac{\ave{(\Delta N_1)^2}}{\ave{N_1}^2} + \frac{\ave{(\Delta N_2)^2}}{\ave{N_2}^2} - \frac{\ave{\Delta N_1 \Delta N_2 } }{\ave{N_1} \ave{N_2}} 
\end{equation}
\begin{equation}
\sigma_{mix}^2 = \frac{1}{\ave{N_1}^2} + \frac{1}{\ave{N_2}^2}
\end{equation}
As we discussed before, provided the chemical parameters do not change across systems, $\Psi_{N_1/N_2}^{N_1} = \ave{N_1} \sigma^{N_1/N_2}_{dyn}$ should be strictly independent of centrality and system size.  This requirement is not satisfied by the SPS scan \cite{na49fluct} since there $\mu/T$ does vary considerably.    It is however satisfied by the RHIC upper energies.  Thus, the Grand-Canonical statistical model predicts a flat dependence with system size at these energies. 
  The canonical model, on the other hand, would predict a ``kink'' within the same centrality as when the strangeness correlation volume approaches the thermodynamic limit \cite{canonical1,westfall}.    Note that a scan in centrality and system size within the same energy should however produce the same approximately horizontal bands in $\Psi$ as are seen at RHIC, since thermal parameters are to a good approximation independent of system size to the lowest centrality \cite{jan_size}: Each value of $\sqrt{s}$, and hence $\mu/T$, should produce a band when scanned in centrality and system size.
\begin{figure}[t]
\begin{center}
\epsfig{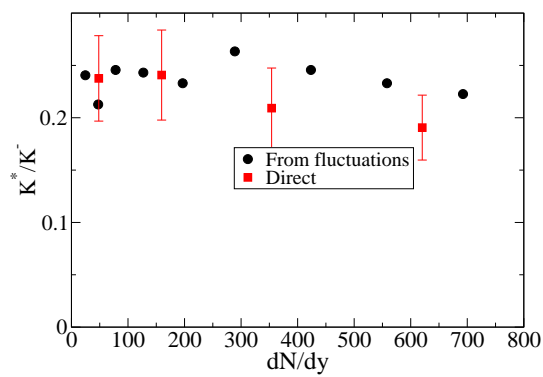}
\caption{\label{fluctreso}   $K^*/K$ implied from fluctuations and directly measured.   Fluctuations and resonances are respectively from \cite{fluctstar,fluctreso}}
\end{center}
\end{figure}
The results can be seen on Fig. \ref{fluctstar}.  None of the models come out perfectly, but the Grand-Canonical model is qualitatively much more similar to the data than the canonical one, as no kink is visible there.
The discrepancy in $\Psi$ between 200 GeV and 62 GeV, the slight upward trend in centrality, and the lack of scaling between A-A and Cu-Cu should however be closely watched, as the slight increase of fluctuations with multiplicity can not be accounted for by {\em any} statistical model, although one has to be careful as it could be an artifact of the event mixing procedure \cite{westfall}.
  
\begin{figure}[h]
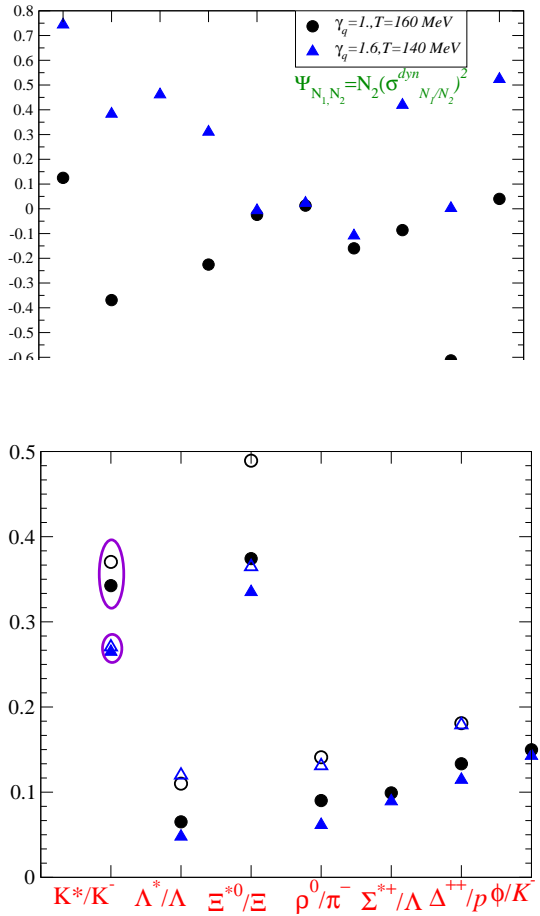

\begin{center}
\epsfig{width=7cm,figure=plotall_fluct.eps}
\epsfig{width=7cm,figure=plotall_reso_fluct.eps}
\caption{\label{plotrhic}  Equilibrium and non-equilibrium  predictions from fluctuations (top panel) and resonances (bottom panel) with the statistical parameters taken from \cite{gammaq_energy}.  In the bottom panel, Full symbols show the $\ave{N^*}/N$ ratio inferred from the correlation in the two models, while empty symbols show, wherever possible, the estimate from $\sigma_{dyn}$ comparisons  }
\end{center}
\end{figure}

Quantitatively, as reported earlier, $\Psi$ is modeled much better with the inclusion of the light-quark non-equilibrium parameter $\gamma_q$, due to Bose-Einstein enhancement of fluctuations \cite{prcfluct}.   It remains to be seen weather the measurement of fluctuations of {\em 
more} particles (Fig. \ref{plotrhic} top panel) will corroborate this conclusion.

We now turn to a potentially important use of fluctuations, the constraint on resonance reinteraction between chemical freeze-out and thermal freeze-out \cite{jeon}.   The former can be estimated, to a good approximation, by comparing the fluctuations of same-charged particles (uncorrelated) and opposite charged particles (correlated by $K^*$) 
\begin{equation}
\frac{\ave{N_{K^{*0}}}}{\ave{N_{K^-}}} \simeq \frac{3}{4} \left( \Psi_{K^-/\pi^-}^{\pi^-} -  \Psi_{K^+/\pi^-}^{\pi^-}  \right)
\end{equation}
Comparing the fluctuation estimate  of $\frac{K^{*0}}{K^-}$ to a direct measurement should yield the amount of $K^{0*}$ destroyed by rescattering or regenerated through pseudo-elastic interactions.

This analysis is shown in Fig. \ref{fluctreso},with resonance values are taken from \cite{fluctreso}.   Rescattering and regeneration have,within error bar little or no effect on the final abundance of $K^*$s.    Either chemical and thermal freeze-out proceed very close to each other, so the amount of reinteraction is negligible, or rescattering and regeneration of detectable resonances cancel each other out to the degree of approximation allowed by the error bars ($10-15 \%$).    This margin appears somewhat below the estimates from transport models, which are $\sim 40\%$ \cite{urqmdreso}, and inline with fits assuming sudden freezeout \cite{jan_search,flork}. 
It would be very interesting to investigate weather a transport model \cite{konch} could be tuned to simultaneously reproduce the fluctuation inference and direct measurement of $\frac{K^{*0}}{K^-}$.

Unfortunately, the baroqueness of the resonance decay tree severely limits  the feasibilness of such graphic methods, as the bottom panel of Fig \ref{plotrhic} shows.     $K,\pi$ is a special pair of particles in that there is only one type of resonances that decays into both, the $K^*$, and its lightest state is considerably lighter than the heavier ones.   No similar definition is possible for $\Sigma^*/\Lambda,\rho/\pi$ and $\phi/K$, since the resonance decays equally into all pairs of decay products.
For other resonances, cross-contamination destroys any value of fluctuations as a {\em graphic} tool of reinteraction time.
This does not mean, however, that such resonances are useless for constraining reinteraction time, since both fluctuations \cite{prcfluct} and resonances \cite{ratio1,ratio2} provide a tight constraint on all statistical models.
If a statistical model consistently describes fluctuations, but over-predicts resonance abundances, it could be taken as an indication of a long reinteraction times with detailed balance.   A casual look at experimental data \cite{fluctreso} shows this does not seem to be the case, since most resonances are {\em under-predicted} by statistical models.

In conclusion, we have provided an overview of the current results in the theoretical interpretation of both resonances and short-lived rations within the statistical model.    We discussed how these observables should scale with different energies and system sizes within various statistical models available on the market.   We concluded that the energy dependence of the ratios of the type $\ave{N^*}/\ave{N}$, as well as a simultaneus description of fluctuations and particle yields, can be very useful for clarifying present theoretical uncertainities.

G.T. acknowledges the financial support received from the Helmholtz International
Center for FAIR within the framework of the LOEWE program
(Landesoffensive zur Entwicklung Wissenschaftlich-\"Okonomischer
Exzellenz) launched by the State of Hesse and the EMMI foundation


\begin{thebibliography}{15}



\bibitem{hornreso} 
  G.~Torrieri and J.~Rafelski,
  Phys.\ Rev.\ C {\bf 75}, 024902 (2007)
  [nucl-th/0608061].


\bibitem{ratio1}
  G.~Torrieri and J.~Rafelski,
  Phys.\ Lett.\  B {\bf 509}, 239 (2001)

\bibitem{ratio2}
  J.~Rafelski, J.~Letessier and G.~Torrieri,
  Phys.\ Rev.\  C {\bf 64}, 054907 (2001)
  [Erratum-ibid.\  C {\bf 65}, 069902 (2002)]


\bibitem{prcfluct}
  G.~Torrieri, S.~Jeon and J.~Rafelski,
  Phys.\ Rev.\  C {\bf 74}, 024901 (2006)

\bibitem{westfall} 
  G.~Torrieri, R.~Bellwied, C.~Markert and G.~Westfall,
  J.\ Phys.\ G G {\bf 37}, 094016 (2010)




\bibitem{history1} 
E.~Fermi
{Prog. Theor. Phys.} {\bf 5}, 570 (1950). 
\bibitem{history2} 
I. Pomeranchuk  
{Proc. USSR Academy of Sciences} (in Russian) 
{\bf 43}, 889  (1951).  
 \bibitem{history3} 
  LD~Landau, 
  Izv.\ Akad.\ Nauk Ser.\ Fiz.\  {\bf 17}  51-64  (1953). 
 \bibitem{history4} 
R. Hagedorn R  
 {Suppl. Nuovo Cimento}  {\bf 2}, 147 (1965). 
 

\bibitem{jansbook}
~Letessier J,  ~Rafelski J (2002),
Hadrons  quark - gluon plasma,
Cambridge Monogr.\ Part.\ Phys.\ Nucl.\ Phys.\ Cosmol.\  {\bf 18}, 1,
and references therein
 

\bibitem{bdm}
  P.~Braun-Munzinger, D.~Magestro, K.~Redlich and J.~Stachel,
  Phys.\ Lett.\  B {\bf 518}, 41 (2001)
 

\bibitem{equil_energy} 
  J.~Cleymans, H.~Oeschler, K.~Redlich and S.~Wheaton, 
  arXiv:hep-ph/0607164. 

\bibitem{becattini} 
  F.~Becattini et. al., 
  Phys.\ Rev.\ C {\bf 69}, 024905 (2004) 


  


\bibitem{nuxu}
  J.~Cleymans, B.~Kampfer, M.~Kaneta, S.~Wheaton and N.~Xu,
  Phys.\ Rev.\  C {\bf 71}, 054901 (2005)

\bibitem{share}
  G.~Torrieri, S.~Steinke, W.~Broniowski, W.~Florkowski, J.~Letessier 
and J.~Rafelski,
  Comput.\ Phys.\ Commun.\  {\bf 167}, 229 (2005)



\bibitem{castorina}
  F.~Becattini, P.~Castorina, A.~Milov and H.~Satz,
  J.\ Phys.\ G G {\bf 38}, 025002 (2011)


\bibitem{jaki}
  J.~Noronha-Hostler, C.~Greiner and I.~A.~Shovkovy,
  Phys.\ Rev.\ Lett.\  {\bf 100}, 252301 (2008)


\bibitem{pbmel}
  K.~Redlich, A.~Andronic, F.~Beutler, P.~Braun-Munzinger and J.~Stachel,
  J.\ Phys.\ G {\bf 36}, 064021 (2009)


\bibitem{sharev2}
  G.~Torrieri, S.~Jeon, J.~Letessier and J.~Rafelski,
  Comput.\ Phys.\ Commun.\  {\bf 175}, 635 (2006)


\bibitem{jeon}
  S.~Jeon and V.~Koch,
  Phys.\ Rev.\ Lett.\  {\bf 83}, 5435 (1999)


\bibitem{shuryak} 
  M.~A.~Stephanov, K.~Rajagopal and E.~V.~Shuryak,
  Phys.\ Rev.\ D {\bf 60}, 114028 (1999)
  [hep-ph/9903292].

\bibitem{canonical1}
  S.~Hamieh, K.~Redlich and A.~Tounsi,
  Phys.\ Lett.\  B {\bf 486}, 61 (2000)


\bibitem{horn} 
  M.~Gazdzicki, M.~Gorenstein and P.~Seyboth,
  Acta Phys.\ Polon.\ B {\bf 42}, 307 (2011)


\bibitem{gammaq_energy} 
  J.~Letessier and J.~Rafelski,
  Eur.\ Phys.\ J.\ A {\bf 35}, 221 (2008)

\bibitem{visc1} 
  G.~Torrieri and I.~Mishustin,
  Phys.\ Rev.\ C {\bf 78}, 021901 (2008)
  [arXiv:0805.0442 [hep-ph]].

\bibitem{visc2} 
  G.~Torrieri, B.~Tomasik and I.~Mishustin,
  Phys.\ Rev.\ C {\bf 77}, 034903 (2008)
  [arXiv:0707.4405 [nucl-th]].


\bibitem{jan_search}
  G.~Torrieri and J.~Rafelski,
  New J.\ Phys.\  {\bf 3}, 12 (2001)


\bibitem{flork} 
  A.~Baran, W.~Broniowski and W.~Florkowski,
  Acta Phys.\ Polon.\ B {\bf 35}, 779 (2004)

\bibitem{urqmdreso} 
  S.~Vogel and M.~Bleicher,
  nucl-th/0505027.


\bibitem{fluctreso}
  J.~Adams {\it et al.}  [STAR Collaboration],
  Phys.\ Rev.\ Lett.\  {\bf 97}, 132301 (2006)

\bibitem{na49reso} 
  V.~Friese [NA49 Collaboration],
  Nucl.\ Phys.\ A {\bf 698}, 487 (2002).


\bibitem{fair} 
  M.~Bleicher, M.~Nahrgang, J.~Steinheimer and P.~Bicudo,
  arXiv:1112.5286 [hep-ph].

\bibitem{rhicscan} 
  G.~Odyniec,
  J.\ Phys.\ G G {\bf 37}, 094028 (2010).

\bibitem{nica}
  A.~N.~Sissakian, A.~S.~Sorin and V.~D.~Toneev,
  Phys.\ Part.\ Nucl.\  {\bf 39} (2008) 1062.


\bibitem{canfluct}
  V.~V.~Begun, M.~Gazdzicki, M.~I.~Gorenstein and O.~S.~Zozulya,
  Phys.\ Rev.\  C {\bf 70}, 034901 (2004)


\bibitem{methods}
  C.~Pruneau, S.~Gavin and S.~Voloshin,
  Phys.\ Rev.\  C {\bf 66}, 044904 (2002)





\bibitem{konch} 
  V.~P.~Konchakovski, S.~Haussler, M.~I.~Gorenstein, E.~L.~Bratkovskaya, M.~Bleicher and H.~Stoecker,
  Phys.\ Rev.\ C {\bf 73}, 034902 (2006)



\bibitem{fluctstar}
        J.~Adams {\it et al.}  [STAR Collaboration],
  Phys.\ Rev.\ Lett.\  {\bf 103}, 92301 (2009)

\bibitem{na49fluct}
  C.~Alt {\it et al.}  [NA49 Collaboration],
  Phys.\ Rev.\  C {\bf 79}, 044910 (2009)



\bibitem{jan_size}
  J.~Rafelski, J.~Letessier and G.~Torrieri,
  Phys.\ Rev.\  C {\bf 72}, 024905 (2005)


\end{thebibliography}
\end{document}